# Predicting structure-dependent Hubbard *U* parameters for assessing hybrid functional-level exchange via machine learning

Zhendong Cao[1,2], Guanghui Cai[1,2], Fankai Xie[1,2], Huaxian Jia[5], Wei Liu[5], Yaxian Wang[1,2],

Feng Liu[6], Xinguo Ren[1,3]*, Sheng Meng[1,2,3]* , Miao Liu[1,3,4]*

[1]*Beijing National Laboratory for Condensed Matter Physics, Institute of Physics, Chinese Academy of Sciences, Beijing 100190, China*

[2]*School of Physical Sciences, University of Chinese Academy of Sciences, Beijing 100190, China*

[3]*Songshan Lake Materials Laboratory, Dongguan, Guangdong 523808, China*

[4]*Center of Materials Science and Optoelectronics Engineering, University of Chinese Academy of Sciences, Beijing 100049, China*

[5]*Tencent AI Lab, Tencent, Shenzhen 518057, China*

[6]*Department of Materials Science and Engineering, University of Utah, Salt Lake City, Utah 84112, USA*

\*Corresponding author:    renxg@iphy.ac.cn, smeng@iphy.ac.cn, mliu@iphy.ac.cn

Z. Cao and G. Cai contributed equally to this work

**Abstract**

DFT+$U$ is a widely used treatment in the density functional theory (DFT) to deal with correlated materials that contain open-shell elements, whereby the quantitative and sometimes even qualitative failures of local and semilocal approximations can be corrected without much computational overhead. However, finding appropriate $U$ parameters for a given system is non-trivial and usually requires computationally intensive and cumbersome first-principles calculations. In this *Letter*, we address this issue by building a machine learning (ML) model to predict material-specific $U$ parameters only from the structural information. A ML model is trained for the Mn-O chemical system by calibrating their DFT+$U$ electronic structures with the hybrid functional results of more than Mn-O 3000 structures. The model allows us to determine a reliable $U$ value (MAE=0.128 eV, $R^2$=0.97) for any given structure at nearly no computational cost; yet the obtained $U$ value is as good as that obtained from the conventional first-principles methods. Further analysis reveals that the $U$ value is primarily determined by the local chemical structure, especially the bond lengths, and this property is well captured by the ML model developed in this work. This concept of the ML $U$ model is universally applicable and can considerably ease the usage of the DFT+$U$ method by providing structure-specific, readily accessible $U$ values.

*Introduction.* – The complex nature of many-body interactions makes it a long-standing challenge to obtain highly accurate exchange-correlation (XC) energy functional for density functional theory (DFT). The semi-local XC functionals, such as the generalized gradient approximation (GGA) of Perdew, Burke, and Ernzerhof (PBE) [1] and many others [2,3], have well-known self-interaction issues, failing to describe the energy bands of many compounds correctly, especially the ionic compounds with open-shell elements. One viable solution is to add intra-atomic interactions between electrons to mitigate the self-interaction error intrinsic to the local or semi-local XC functionals, namely the Hubbard $U$ correction. The DFT+$U$ method, which was first proposed by Anisimov et al. [4] and further developed by Dudarev et al. [5], introduces an on-site Coulomb interaction term to the energy functional to penalize for the partial occupation of the localized orbitals and can correctly predict certain behaviors of strongly correlated systems, e.g., Mott insulators [6], without significantly increasing the computational cost. However, finding appropriate $U$ values for a given material system is not an easy task. Previously, Wang et al. fitted the $U$ for transition metal oxides according to the experimental chemical reaction enthalpy, and a set of recommended $U$ values were obtained for open-shell metal species [7]. As a result, a series of *ad hoc* energy corrections are needed to systematically tame the reaction enthalpy to match experiments [8]. First-principles approaches, such as the linear response method [9] and constrained random phase approximation (cRPA) [10–12], were invented to determine the $U$ value for a given system, however, these approaches typically increase the computational cost by at least 10 times. Yu et al., later on, employed the machine learning (ML) method, specifically the Bayesian optimization (BO), to extract the $U$ value according to higher-level ab initio results [13]. Such a method has been successfully applied to interface [14,15] and superlattice [16] systems, but the computational overhead is still significant in order to generate the training dataset.

One of the major challenges to find the generic $U$ values is that the Hubbard $U$ correction introduces short-range interactions between electrons, and the $U$ value is significantly influenced by the local charge density of the system and is system-and orbital-dependent. Therefore the $U$ value for one system is unlikely to be transferable to another system, especially when these two systems are far apart from each other in terms of their electronic structures. For example, it was found that the effective $U$ value is 1.5 eV in $Ce_2O_3$ [17], but drastically increased to 6.3 eV in $CeO_2$ [18]. It has been demonstrated that the Hubbard $U$ is crucially structure-dependent when the system undergoes pressure-induced phase transition [19], or high-spin to low-spin magnetic phase transition [20] as the pressure and spin polarization indeed modulate the structures of those systems. It thus calls for the appropriate choice of $U$ values to obtain a physically more accurate potential energy surface [21]. Hence, establishing the underlying connections between the Hubbard $U$ parameters of a certain element and its local structure, oxidation state, as well as its orbital hybridization, is the prerequisite for building a predictive model to foretell the $U$ value when a structure is given.

In this *Letter,* we tackle this problem by employing our in-house high-throughput BO-based workflow, and using the Mn-O compounds as our model system. We employ more than 3000 Mn-O configurations whose band gaps and energy bands are fitted to the high-level hybrid functional result (Heyd-Scuseria-Ernzerhof functional [22,23], HSE) as for the abundant structure availability of the Mn-O system. This step essentially follows what has been done in Ref. [13]. In a second, and more important step, we carry on to employ a supervised random forest ML algorithm [24,25] to train a predictive Hubbard $U$ model, which then predicts the $U$ value with sufficient accuracy and efficiency for any given Mn-O structure, even those not included in the training dataset. The obtained ML model shows a remarkable accuracy and reaches the coefficients of determination $R^2$=0.97 and mean absolute error (MAE) of 0.128 eV for $U$ values. More importantly, it is unraveled by the regression that the $U$ value is primarily associated with the bond length, which is consistent with the cRPA theory.

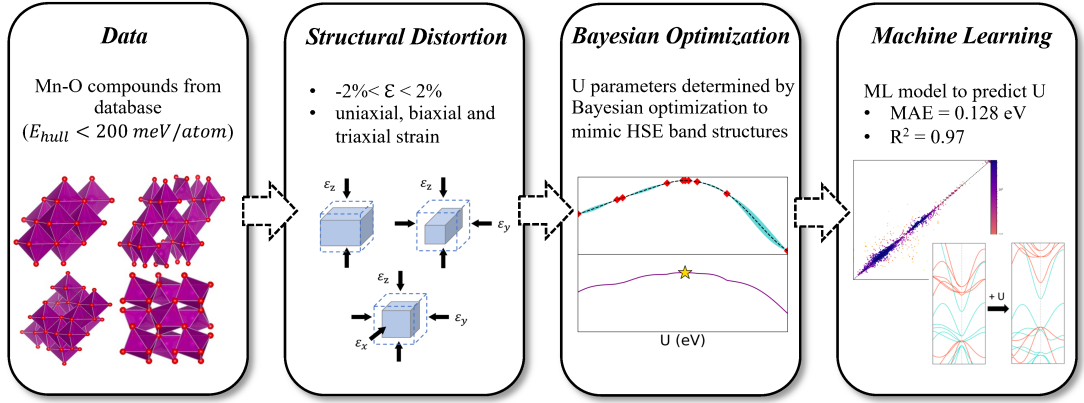

FIG. 1. The flow chart of the machine learning process to create a structure-dependent Hubbard $U$ model.

*Workflow.* – The process including data generation, structural distortion, BO, and ML is schematically illustrated in Fig. 1. The thermodynamic stability of the compounds is evaluated by the physical quantity of energy above hull ($E_{hull}$), which is the reaction enthalpy required to decompose a material to other stable compounds [26,27]. We start from selecting the thermodynamically stable ($E_{hull} < 200 meV/atom$) and Mn-O chemical systems with reasonable size ($N_{atom} < 20$) from 312 possible structures generated from the Atomly [28] materials database and end up with 67 individual compounds. To enlarge the size of our structural space, we apply uniaxial, biaxial, and triaxial strain of $-2\% < \varepsilon < 2\%$ to the selected 67 compounds and obtain 3724 Mn-O structures. The strain deformation step is employed to modify the screened local Coulomb interaction $U$ [29,30] and ensures that our model surveys among a large enough structural space to yield a good model extrapolation. The BO method, which is developed by Yu et al., is employed to determine the Hubbard $U$ parameter within the PBE+$U$ method by fitting to the HSE band gap and band structures for all the structures. In the present work, the PBE+$U$ calculations are carried out by VASP [31] and the HSE band structures are computed using the FHI-aims code [32–

34]. In this way, a $U$ value for Mn $3d$ state can be fitted for each structure. We note that materials were modeled as ferrimagnetic ordering. In principle, higher-level methods, such as the GW [35] or coupled cluster singles and doubles (CCSD) [36]-level of the calculation, can be used too if one has enough computational resources, but we use the HSE as the "ground truth" for its viable efficiency. Finally, a ML model is constructed by harnessing the Random Forest Regression (RFR) to directly predict the optimal Hubbard $U$ parameters for any Mn-O structures. The computational details can be found in Supplemental Material.

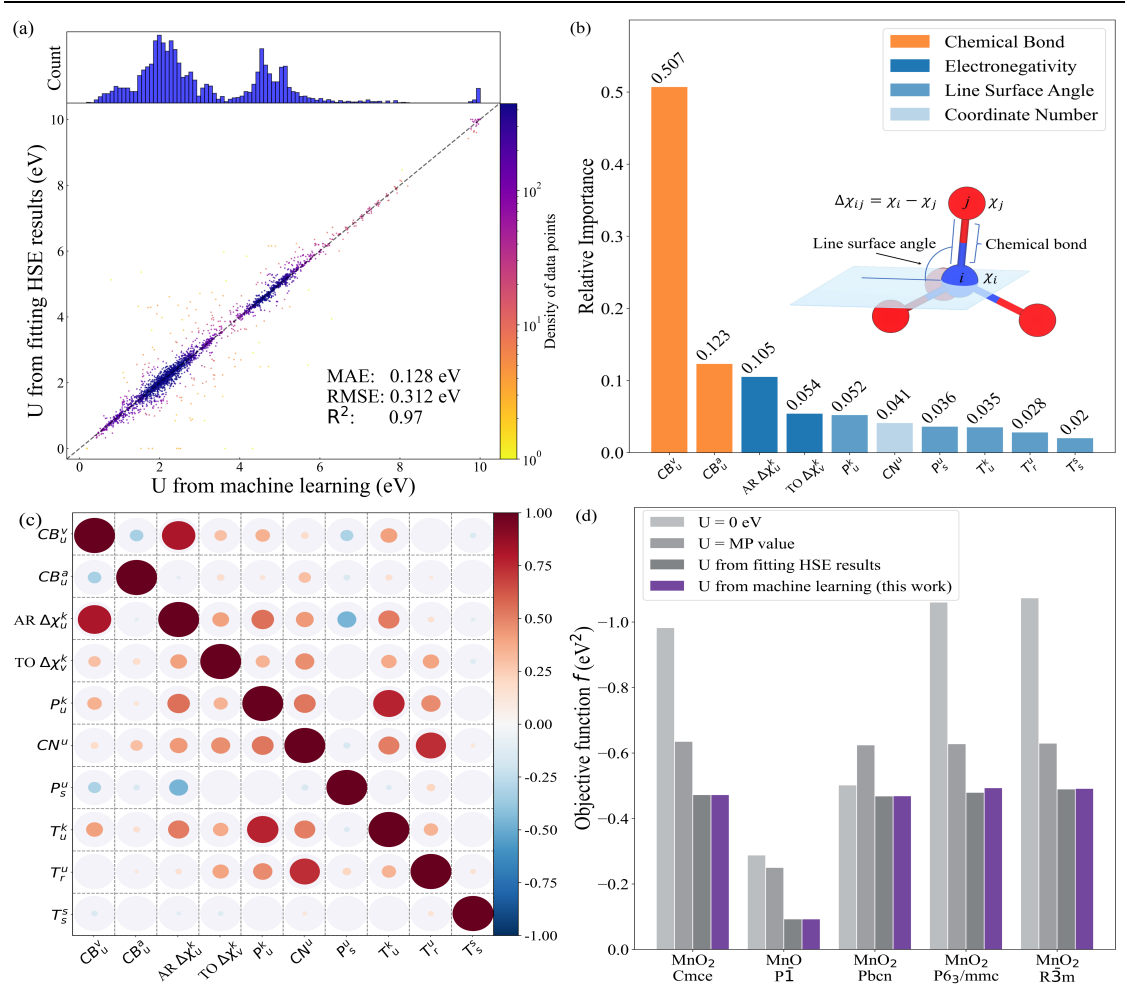

FIG. 2. (a) The comparison of Hubbard $U$ values from fitting the HSE results (Bayesian optimization) and machine learning. (b) The importance ranking of all descriptors in prediction, highlighting the most significant factor from the bond length. Different kinds of descriptors are colored in different colors, as depicted in the legend. The superscript and subscript of the descriptor denote the intersite and intra-site mathematical operation, respectively (see Supplemental Material). (c) Pearson correlation coefficient matrix of all descriptors. The radii of circles represent the absolute magnitude of coefficients. (d) The performance of different methods to determine the Hubbard $U$. The objective function is employed as the evaluation of the performance of different methods. Here five structures are chosen from the Atomly database with their chemical formulas and space groups listed by the horizontal axis.

*Model validation.* − Random forests are a combination of decision trees that individually make predictions on each input and the overall prediction determined by a majority voting process. We evaluate the prediction ability of our RFR model by plotting the out-of-bag (OOB) error, which can be analogous to the conventional cross-validation error but provides a global error estimated for all data points, shown in Fig. 2(a) along with the detailed data distribution. Our model can predict the Hubbard $U$ values fairly accurately with MAE=0.128 eV and $R^2$=0.97, meaning that the predicted $U$ value falls into a small error range of ±0.128 eV statistically. The distribution of Hubbard $U$ values shows two peaks due to the uneven distribution of Mn-O bond length which will be discussed later in the paper. It is noticeable that for some structures, their electronic structures are insensitive to the change of Hubbard $U$ parameters, and hence the heavily-structure-dependent model does not apply so well to these compounds, causing the deficiency of the ML model to some extent, but the model overall has good accuracy. To gain a better insight into the physical connection between the Hubbard $U$ parameters and the materials' properties, we sort out the 10 most important descriptors for the $U$ value prediction, as shown in Fig. 2(b). There are *CB*, $\Delta\chi$, *P*, *T*, and *CN* (see Supplemental Material, Table S1 for their definitions). Further, these descriptors can be divided into four categories: chemical bond (*CB*), electronegativity difference ($\Delta\chi$), line surface angle (*P*, *T*), and coordinate number (*CN*). It is obvious that, other than the electronegativity, nearly all the decisive descriptors, the CB, line surface angle, and coordinate number, are primarily a function of the atomistic structure of a compound, indicating that the $U$ is primarily a structure-dependent parameter. The CB length is the most important factor for the predictions (50.7%), which is computed as the

(normalized) total reduction of the criterion brought by that feature.

Upon categorizing these 10 factors into four aspects, these parameters have a close connection with each other. Therefore, we look into the Pearson correlation matrix (Fig. 2(c)) of the 10 descriptors. It can be seen that the choice of the descriptors is fairly orthogonal as those descriptors are weakly coupled with each other, except for the CB descriptor ($CB_u^v$) and the electronegativity descriptor (AR $\Delta\chi_u^k$, AR represents Allred-Rockow electronegativity). This is expected considering that the electronegativity descriptors themselves are essentially derived from the crystal structure and the atom species. On the other hand, the strength of a CB should depend on the electronegativity difference ($\Delta\chi$) between the two bonding atoms. The $\Delta\chi$ between atoms bonded together will greatly affect the charge density distribution of the local structure, thereby influencing the local electronic screening. Therefore, the correlation between electronegativity and the CB length further suggests it is viable to assign a $U$ value to a compound based on its structure.

Fig. 2(d) presents the performance of the ML methods in comparison with the direct energy calculation with and without corrections. The objective function to evaluate the performance of a given $U$ value is defined as [13]:

$$f(\vec{U}) = -\alpha_1\left(E_g^{\text{HSE}} - E_g^{\text{PBE}+\text{U}}\right)^2 - \alpha_2(\Delta\text{Band})^2. \quad (1.)$$

Here, $\vec{U} = [U^1, U^2, ..., U^n]$ is the vector of $U$ values applied to different atomic species. $E_g^{\text{HSE}}$ and $E_g^{PBE+U}$ represent the band gaps calculated by the HSE and PBE+$U$ functionals. $\Delta$Band is defined as the mean squared deviation of the PBE+$U$ band structures with respect to their HSE counterparts, similar to Ref. [37]:

$$\Delta\text{Band} = \sqrt{\frac{1}{N_E}\sum_{i=1}^{N_k}\sum_{j=1}^{N_b}\left(\epsilon_{\text{HSE}}^j[k_i] - \epsilon_{\text{PBE}+\text{U}}^j[k_i]\right)^2}, \quad (2.)$$

where $N_E$ represents the total number of eigenvalues $\epsilon$. The summation goes through $N_k$ $K$-points and $N_b$ selected bands, and obviously $N_E = N_k * N_b$. The coefficients $\alpha_1$ and $\alpha_2$ are the control parameters that assign different weights to the band gaps and band structures. We set $\alpha_1 = 0.25$ and $\alpha_2 = 0.75$ as default in agreement with Ref. [13]. The closer the objective function is to 0, the better the PBE+$U$ calculations reproduce the HSE results. It can be observed that our model outperforms the traditional treatment which used a fixed $U$ value for a given compound and reaches the same accuracy as that of the BO method developed by Yu et al. [13]. For all data points, the MAE difference of the objective function between ours and that of Yu et al. is about $0.01 \text{eV}^2$ (Supplemental Material, Fig. S1). Our method can predict the $U$ value and reproduces the band gap and band structures obtained from HSE without running the $U$ parameter calculations, thus making it easier and more efficient for performing DFT+$U$ calculations for strongly correlated systems.

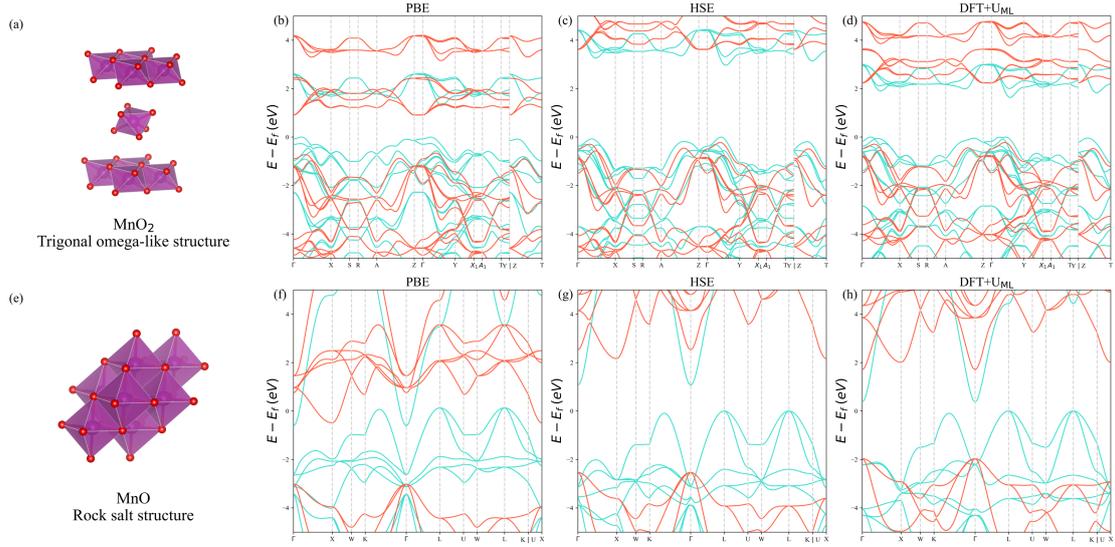

FIG. 3. (a) The crystal structure of $MnO_2$. Band structures of $MnO_2$ obtained in different methods: (b) PBE; (c) HSE; (d) PBE with ML predicted $U$. (e) The crystal structure of MnO. Band structures of MnO were obtained in different methods: (f) PBE; (g) HSE; (h) PBE with ML predicted $U$. The red line and the cyan line represent the spin-up and spin-down bandstructure, respectively and materials were modeled as ferrimagnetic ordering.

Fig. 3 demonstrates the performance of PBE with predicted $U$ for $MnO_2$ and MnO compounds. It can be found that the PBE functional is unable to fully capture the electronic structure of the Mn-O compounds (and in fact also other transition metal oxides) due to the incomplete self-interaction error cancellation [38–41], and yields an overestimation of Coulomb repulsion. For $MnO_2$ (Fig. 3(a)), the PBE band gap of 0.91 eV (Fig. 3(b)) is considerably underestimated compared to the HSE result of 2.96 eV (Fig. 3(c)). Moreover, the locations of the conduction band minimum (CBM) and valence band maximum (VBM), as well as the spin-up and spin-down channels, from PBE are drastically different from those from HSE. Using the Hubbard $U$ predicted in this work, we obtain the spin-polarized band structure that matches the HSE result well (Fig. 3(d)). For MnO, the usage of our predicted Hubbard $U$ successfully corrects the band position closer to the HSE values, and more importantly results in a band gap opening for this compound. This material is a conductor according to PBE (Fig. 3(f)) and a semiconductor with a band gap of 0.39 eV in PBE+$U$ (Fig. 3(g)), amending the PBE result significantly. We note that adding a proper $U$ correction still underestimates the band gap to some extent, e.g. in $MnO_2$, the ML $U$ correction increases the gap to 1.63 eV, whereas the HSE band gap is 2.96 eV; in MnO, the ML $U$ correction increases the gap to 0.39 eV, whereas the HSE band gap is 1.09 eV (Fig. 3(h)). It reflects that the Hubbard $U$ correction cannot completely capture the features of exchange interactions in HSE. Overall, our ML model predicts reliable Hubbard $U$ values that apply well to the PBE+$U$ calculations and reproduces the qualitative features of HSE band structures.

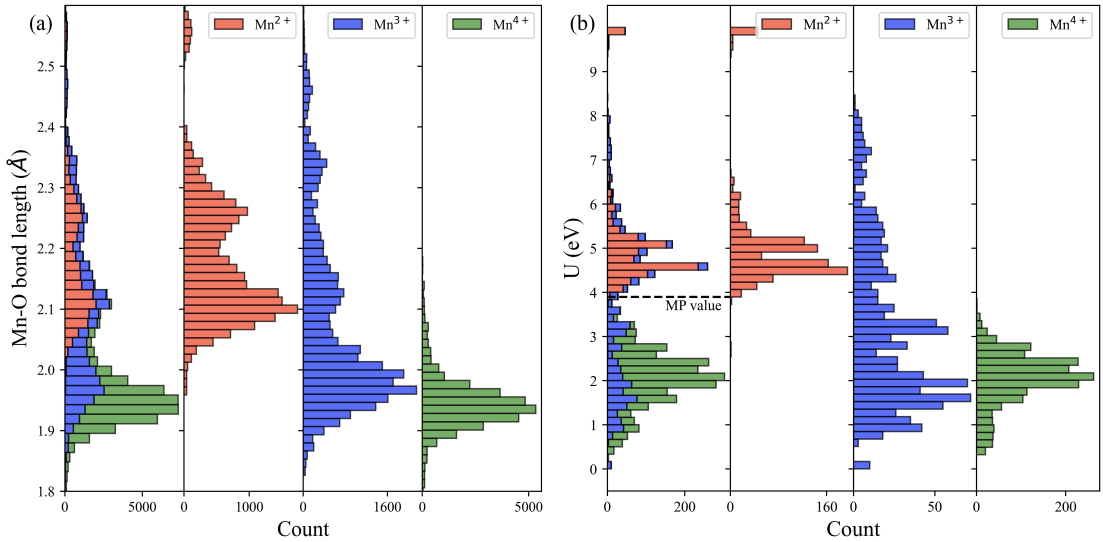

FIG. 4. The (a) bond length and (b) Hubbard $U$ distributions with the contribution of different valence states ($Mn^{2+}$, $Mn^{3+}$, and $Mn^{4+}$) are colored in red, blue, and green, respectively. The dashed line denotes the $U$ value used in the Materials Project.

*Structural dependence.* − As discussed in the previous session, our ML model indicates that the Hubbard $U$ parameter is greatly structure-dependent. In order to investigate the correlation between the Hubbard $U$ parameter and Mn-O bond length, their distributions among different valence states are plotted in Fig. 4. It can be seen that the distribution of Mn-O band length is shape-wise similar to that of the $U$ parameters. For example, $Mn^{3+}$ exhibits a wide distribution of bond lengths, while its Hubbard $U$ spreads over a wider range (from 0 to 10 eV) compared to the $Mn^{2+}$ and $Mn^{4+}$ cases. Furthermore, the overall larger Mn-O bond lengths in $Mn^{2+}$ compounds correspond to their overall larger $U$ values, whereas the smaller Mn-O bond lengths in $Mn^{4+}$ compounds lead to smaller $U$ values. The dashed line in Fig. 4(b) represents the single $U$ value used in the Materials Project (MP) [42] as obtained by fitting experimental data, showing an apparent discrepancy with the distribution of Hubbard $U$ values from this work. Given that, it implies that the Mn-O bond length is a sensible parameter to describe the variation in $U$, suggesting the necessity of further investigating the relationship between the Mn-O bond length and the Hubbard $U$.

The Hubbard $U$ parameter can be calculated by the cRPA method as follows:

$$U_{Im,Im'} = \int dr dr' |\phi_{I,m}^{3d}(r)|^2 \widetilde{W}(r,r') |\phi_{I,m'}^{3d}(r')|^2, \tag{3.}$$

where $\phi_{I,m}^{3d}(r)$ is the wavefunction of a Mn 3$d$ orbital at the site $I$ with magnetic angular momentum $m$. $\widetilde{W} = \epsilon^{-1} v = (1 - v\widetilde{X_0})^{-1} v$ denotes the effective screened Coulomb interaction with $v$ denoting the bare Coulomb interaction. $\widetilde{X_0}$ is the polarization function, and its magnitude depends on the electronic response property of the system, which is in turn governed by the atomic structure, in particular the bond lengths between neighboring atoms. Within the cRPA scheme, one needs to compute the microscopic polarization function and screened Coulomb interaction for a given atomic structure, which is *de facto* much more costly than the DFT+$U$ calculation itself. Here, our model directly delivers $U$ values from the atomic structure, circumventing the cumbersome step of calculating the microscopic polarization function, yet capturing the same essential physics, namely, the $U$ value is ultimately determined by the local chemical environment.

*Discussions.* − This work showcases a ML model for predicting Hubbard $U$ to skip the expensive first-principles $U$ value calculation process without sacrificing accuracy. Although the Mn-O system is selected as the model system, the out-of-box models can be created for the community for all the open-shell elements as the $U$ value is essentially local-structure dependent, which is consistent with the findings reported in previous work [21,43]. This method has the advantage that the model allows one to assign an appropriate Hubbard $U$ parameter to a system prior to the DFT calculation and yields improved results that are close to the higher-level methods such as HSE or GW. When applying the DFT+$U$ method to structural relaxations or molecular dynamics simulations, it would be ideal to adjust the $U$ parameter to appropriate values on the fly as the structure evolves [19,44,45]. However, this will become prohibitively expensive if the $U$ value is determined using the conventional first-principles approaches, such as the linear response or the cRPA schemes. The pre-trained ML model, as demonstrated in the present work, will make all this readily happen.

Another advantage is that this approach can be extended to several other properties of systems other than the energy band difference. For example, the model can also calibrate the adhesive energy of the system by including the energies in the objective function. Also, the intersite interaction parameter, $V$ [46], can be also incorporated into the model to further improve its predictive power, which we hope to spark a future investigation.

Moreover, the accuracy and robustness of our ML model can be further enhanced with the reinforced dataset. The purpose of this paper is to showcase this approach, while we are aware that with the hybrid-functional-level treatment adopted, only a small dataset is produced (3724 data points) due to computational cost. If the size of the dataset is presumably enlarged by one or two orders of magnitude, the ML model could evolve into a deep neural network, meanwhile, the model accuracy, extrapolation, and generalization can be greatly enhanced.

Finally, our work demonstrates that the Hubbard $U$ parameter is local-structure dependent to some extent. However, the $U$ value we used in this work is a kind of global measure of the electronic screening effect, which may be not sensitive to the change in local structure. One solution can be to assign the $U$ values for every inequivalent site, which requires a tremendous amount of calculation resources and a fairly large dataset for model training. To this end, we hope efforts can be made by the entire community to collaboratively carry forward this method to generally reliable and efficient models to predict Hubbard $U$ values for all open shell elements.

*Conclusions.* −In summary, we developed a data-driven method for predicting the value of Hubbard $U$ for DFT calculations. Specifically, a ML model is constructed to predict the Hubbard $U$ for Mn-O systems, which can accurately assess the $U$ value of a system without running costly first-principles calculations. It is also demonstrated that the predicted Hubbard $U$ can reproduce hybrid functional-level band gap and band structures without actual hybrid functional-level runs. In addition, our ML model reveals the bond length which shares similar distribution with the Hubbard $U$ is the most decisive factor in determining the $U$ value, which can be justified by cRPA theory. Developing a ML model that can accurately yield appropriate $U$ values for a given structure, without actually running expensive and sophisticated electronic-structure calculations, is a long-sought goal. We demonstrate in this work that this is indeed possible, at least in a given type of system. More work is needed to extend the present model from Mn-O systems to general atomic species and structures, but we don't expect essential difficulties that prevent us from eventually achieving this goal. Our ML model not only opens up a new avenue to calculate Hubbard $U$ values for all open-shell elements, but also provides insights into the physical correlation between the $U$ parameters and local structure in condensed matter, which is relevant to many other important physical questions, such as metal-insulator transitions, superconductivity, magnetic phase transition, etc.

*Acknowledgments.* −We especially thank the Atomly database for data sharing. This research is supported by the National Key R&D Program of China (2021YFA0718700). The computational resource is provided by the Platform for Data-Driven Computational Materials Discovery of the Songshan Lake laboratory. We would also acknowledge the financial support from the Chinese

Academy of Sciences (Grant No. ZDBS-LY-SLH007, XDB33020000, and CAS-WX2021PY-0102) and the National Natural Science Foundation of China (Grand No. 12134012 and 12188101).